# Mapping the intrinsic photocurrent streamlines through micromagnetic heterostructure devices

Morgan Mayes[1,2], Maxwell Grossnickle[1,2], Mark Lohmann[1], Mohammed Aldosary[1], Junxue Li[1], Vivek Aji[1], Jing Shi[1], Justin C.W. Song[3,4], and Nathaniel M. Gabor[1,2*]

[1]*Department of Physics and Astronomy, University of California, 900 University Avenue, Riverside, California 92521, United States.*

[2]*Laboratory of Quantum Materials Optoelectronics, Materials Science and Engineering Building, Room 179, University of California, Riverside, California 92521, United States.*

[3]*School of Physical and Mathematical Sciences Division of Physics and Applied Physics, Nanyang Technological University, Singapore 637371.*

[4]*Institute of High Performance Computing, Agency of Science, Technology and Research, Singapore 138632.*
**Correspondence to: nathaniel.gabor@ucr.edu.*

Like air flowing over a wing, optimizing the flow of electronic charge is essential to the operation of nanoscale devices. Unfortunately, the delicate interplay of charge, spin, and heat in complex devices has precluded detailed imaging of charge flow. Here, we report on the visualization of intrinsic charge current streamlines through yttrium iron garnet micromagnetic heterostructures. Scanning photovoltage microscopy of precisely designed devices leads to striking spatial patterns, with prominent photovoltage features emerging in corners and narrow constrictions. These patterns, which evolve continuously with rotation of an external magnetic field, enable rich spatial mapping of fluid-like flow. Taking inspiration from aerodynamic Clark Y airfoils, we engineer micromagnetic wing shaped devices, called electrofoils, which allow us to precisely contort, compress and decompress flowlines of electronic charge.

At microscopic scales, individual molecules moving through a wind tunnel undergo seemingly random motion. Nevertheless, the macroscopic movement of air around obstacles is governed by collective fluid flow. In an analogous way, we often think of electronic charge transport in terms of diffusion of individual particles, yet current-voltage characteristics in a device are governed by the collective flow of current along streamlines. Like the patterns of air flowing in a wind tunnel, the spatial pattern of charge current flux lines gives the density and direction of flow over space and time. This spatial pattern, which depends on both local properties (conductivity) and global boundaries of the system, underpins our description of current-voltage characteristics. Indeed, charge current flow patterns have successfully explained current hot spots close to the corners of two-terminal devices under a magnetic field[1], as well as the drop-off of ohmic non-local voltage in a van der Pauw geometry[2].

Despite being at the core of our understanding of charge transport, the intrinsic global pattern of current flux lines through an electronic device have never been imaged. The main difficulty lies in the fact that conventional transport measurements are unable to probe local electron flow, instead giving access to largescale quantities such as current and voltage. Several techniques have recently emerged to overcome this hurdle, including scanning single electron transistors[3,4], magnetometers[5-8], and quantum gases[9]. Yet, these measurements are limited when investigating buried interfaces or complex magnetic environments, such as those relevant to spin current and thermospintronic flows[10-14]. Even with experimental access to local charge current, probing streamlines requires precisely controlled directional current flow. Such directional flow could be engineered using built-in p-n junctions or material interfaces, yet these inhomogeneities may often mask the natural flow of electronic charge.

Here, we unveil an optoelectronic imaging method to measure the streamlines along which current naturally flows through ultrathin electronic devices. Combining data-intensive scanning photovoltage microscopy with a highly uniform rotating magnetic field, we probe the photoresponse of ultra-high quality micromagnetic heterostructures. Much like tracers in wind tunnels are used to map the flow of air around an aerodynamic airfoil, we use a scanning laser beam as a source of directional charge current to map the flow around precisely engineered wing shaped devices, or electrofoils. Our current flux imaging technique reveals that streamlines can be contorted, compressed or decompressed by changing the shape and angle of attack of the electrofoil devices, in direct analogy to aerodynamic flow. While the spatial pattern of current streamlines is often assumed to be a convenient mathematical abstraction, we show for the first time that current flux lines can be imaged, manipulated, and engineered to control global device properties.

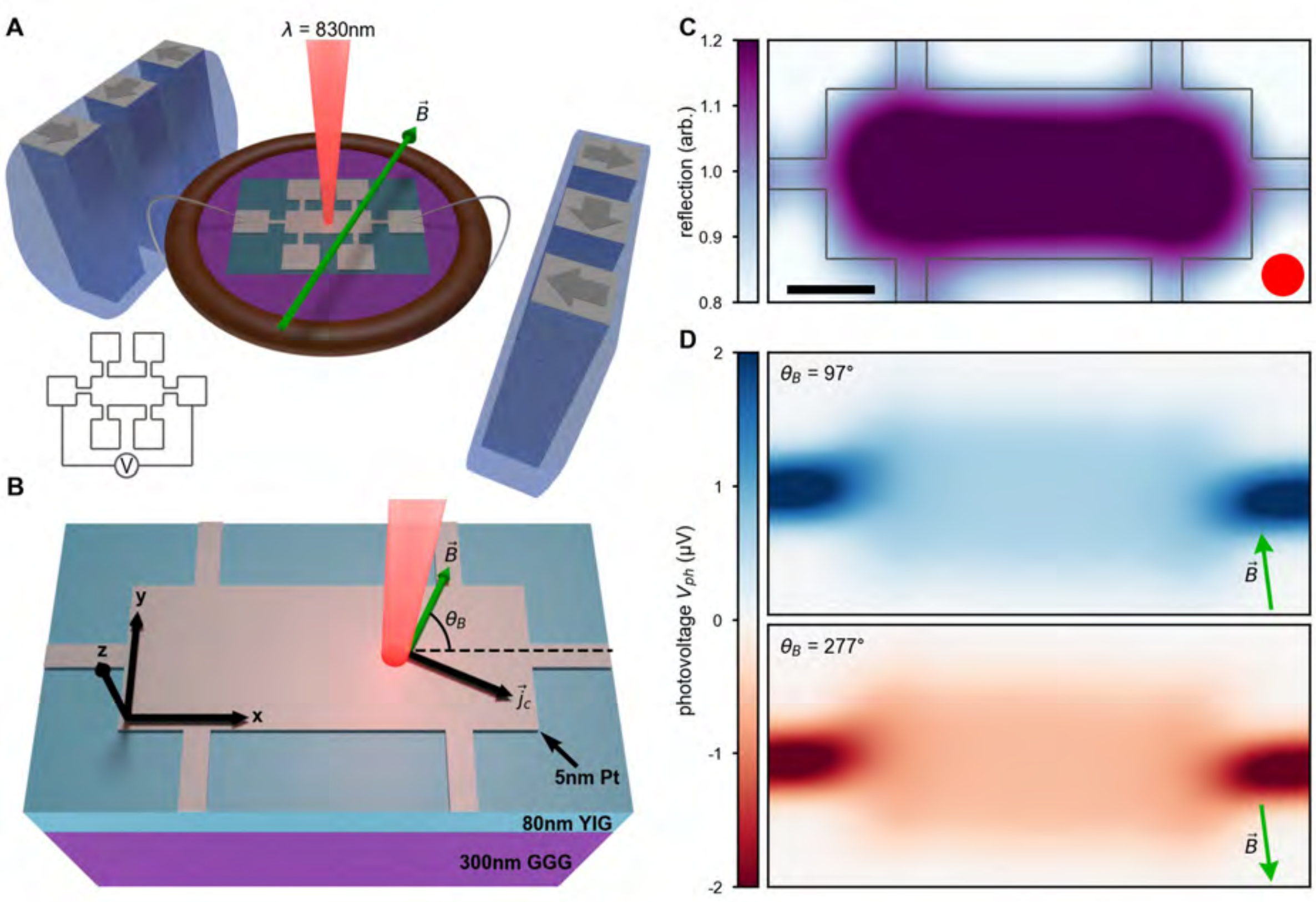

**Fig. 1.** *Current flux microscopy of micromagnetic heterostructure devices.* ***A***, Schematic of the magnetic heterostructure Hall bar device and measurement. Green arrow indicates magnetic field direction on the device. Inset, two-terminal photovoltage configuration, all side contacts are grounded. ***B***, Detailed view of Device 1, a conventional Hall bar defined in ultrathin Pt (5 nm thick) on gadolinium gallium garnet (GGG, 300 nm) and ferrimagnetic layer yttrium iron garnet (YIG, 80 nm). Under laser illumination ($\lambda$ = 830 nm, average power density $1.0 \times 10^4$ W/cm$^2$), thermospintronic response produces a charge current $j_c$ perpendicular to the applied magnetic field *B*. $\theta_B$ is the angle between the magnetic field and the axis defined by the photovoltage probes (*x*-axis). ***C***, Reflection map of Device 1. Red circle indicates the full width at half maximum (FWHM) of the beam spot. Scale bar 50 microns. ***D***, Magnetic field-dependent photovoltage $V_{ph}$ maps at two magnetic field angles $\theta_B$= 97°, 277°, *B*-field direction indicated by green arrows.

Figure 1 shows the scanning magneto-photovoltage microscopy technique used to study ultrathin micromagnetic devices. Using a precisely structured Halbach array of permanent magnets, we establish a magnetic field $\vec{B}$ that can be rotated through the entire solid angle of three-dimensional space (Figure 1A). In the uniform region of the *B*-field, a scanning laser (wavelength $\lambda$ = 830 nm) is used to generate photo-induced voltage response within the metal-magnet heterostructure devices. The devices studied in this work, shown schematically in Figure 1B, are composed of ultrathin platinum patterned on a ferrimagnetic insulator yttrium iron garnet (YIG) thin film, which has been epitaxially grown on gadolinium gallium garnet (supplementary Section S1). As the laser scans over the device, we measure the photovoltage at each point, while simultaneously imaging the device using the back reflected light intensity. Taking advantage of very high image stability, we acquire ~$10^5$ photovoltage measurements, varying in space and magnetic field orientation (supplementary Section S2).

We first characterized a Pt/YIG Hall bar device by measuring photovoltage as a function of laser position and in-plane magnetic field angle $\theta_B$. At a fixed $\theta_B$, we scan the laser across the device and generate maps of the reflected intensity (Figure 1C) and magnetic field-dependent photovoltage $V_{ph}$ (Figure 1D) (supplementary Section S3 and S4). As shown in Figure 1D, top, when we set the magnetic field to be situated across the device ($\theta_B$ = 97$^0$) relative to the voltage probes (positioned along the x-axis labeled in Figure 1B), we observe spatially uniform positive photovoltage that is enhanced along the narrow contacts. When the magnetic field is rotated by 180$^0$, the polarity of $V_{ph}$ changes from positive to negative, yet exhibits otherwise similar spatial features (Figure 1D, bottom).

The photovoltage images of Figure 1D are readily described by a thermospintronic response. Under laser illumination, local charge current in the ultrathin Pt layer is given by:

$$\vec{j}_c = \theta_{SH}\left(2e/\hbar\right)\vec{j}_s \times \vec{\sigma} \quad (1)$$

where $\theta_{SH}$ is the spin Hall angle for platinum, $e$ is the elementary charge, and $\hbar$ is Planck's constant[10,15-21]. The local charge current $\vec{j}_c$ flows in the direction orthogonal to both the spin flow direction $\vec{j}_s$ and the spin polarization $\vec{\sigma}$, each of which is controlled by experiment. The spin polarization vector $\vec{\sigma}$ is parallel to the YIG magnetization, aligning to the applied *B*-field. The spin flow direction $\vec{j}_s$ is anti-parallel to the temperature gradient $\nabla T$ across the Pt/YIG interface and points into the plane of the device under laser heating (supplementary Section S3). In Figure 1D, laser illumination produces a charge current density $\vec{j}_c \propto -\nabla T \times \vec{B}$ along the *x*-axis toward the current-carrying contacts. This charge current in turn gives rise to a photo-induced voltage across the device.

Equation 1 indicates an opportune experimental tool: Using the laser beam as a source of local spin current, we control the direction of local charge current $\vec{j}_c$ by changing the magnetic field angle $\theta_B$ (schematic Figure 1B). Indeed, as shown in Figure 2, optoelectronic measurements taken as $\theta_B$ was varied exhibit striking spatial patterns, with prominent photovoltage features emerging in corners and at narrow constrictions. These patterns are in sharp contrast to measurements taken with the in-plane magnetic field situated across the device, Figure 1D. When the *B*-field is parallel to the voltage probe direction ($\theta_B$ = 177$^0$, Figure 2A middle image), $V_{ph}$ is suppressed in the central region of the device, as expected from ordinary thermospintronic response. However, new spatial features appear at sharp corners and along transverse contacts. As the magnetic field rotates through 180$^0$, the polarity of these anomalous features switches and depends on whether the *B*-field is parallel or anti-parallel to the *x*-axis.

From scanning magneto-photovoltage images, we can examine in detail the evolution of the anomalous photovoltage features as a function of $\theta_B$. Figure 2B shows a rescaled $V_{ph}$ map measured at $\theta_B$ = 357$^0$. For a point in the central region of the device (yellow circle Figure 2B), we plot $V_{ph}$ vs. $\theta_B$ in Figure 2C. We find that the data is well fit by a sinusoidal function $V_{ph} = V_m \cos(\theta_B - \theta_B^{max})$ characterized by two parameters: the photovoltage amplitude $V_m$ and the magnetic field angle at which $V_{ph}$ reaches a maximum, $\theta_B^{max}$. Consistent with thermospintronic response, the photovoltage in the central region (yellow data Figure 2C) reaches a maximum of $V_m$ = 0.6 μV when $\theta_B$ is at a right angle to the *x*-axis ($\theta_B^{max}$ = 90$^0$).

At different positions within the micromagnetic device, we find that maximum photovoltage results from a unique alignment of the magnetic field. If the laser is fixed at the red diamond at top-right in Figure 2B, $V_{ph}$ reaches a maximum when $\theta_B$ = 0$^0$. This is in direct contrast to the blue triangle at bottom-right in Figure 2B, for which the photovoltage reaches a maximum when $\theta_B$ = 180$^0$. When compared to the central region, both spatial features exhibit sinusoidal behavior that is offset by a 90$^0$ phase (Figure 2C). To better visualize the variations of amplitude and phase patterns, we generate images of the photovoltage amplitude $V_m(x,y)$ and angular offset $\theta_B^{max}(x,y)$ at *all* points in space using the sinusoidal fits in Figure 2C. While the amplitude $V_m(x,y)$ (Figure 2D) looks qualitatively similar to the ordinary photoresponse of Figure 1, the angular offset $\theta_B^{max}(x,y)$ exhibits a rich structure that gives the magnetic field angle required to maximize the photovoltage (Figure 2E).

As we now explain, this phase map underpins a rich spatial pattern of fluid-like flowlines through the device. To see this, in Figure 3A, we superimpose a two-dimensional vector field over the image of $\theta_B^{max}(x,y)$. Green arrows indicate the direction of the *B*-field that yields maximum photovoltage amplitude. Crucially, since *B*-field controls the direction of the local thermospintronic response (Equation 1), we plot black arrows that indicate the direction of the local thermospintronic current density that yields the maximum global photovoltage measured. By interpolating the black local current density arrows in Figure 3A, we obtain a flow field through all points in the device, shown in Figure 3B (see also supplementary Section S4).

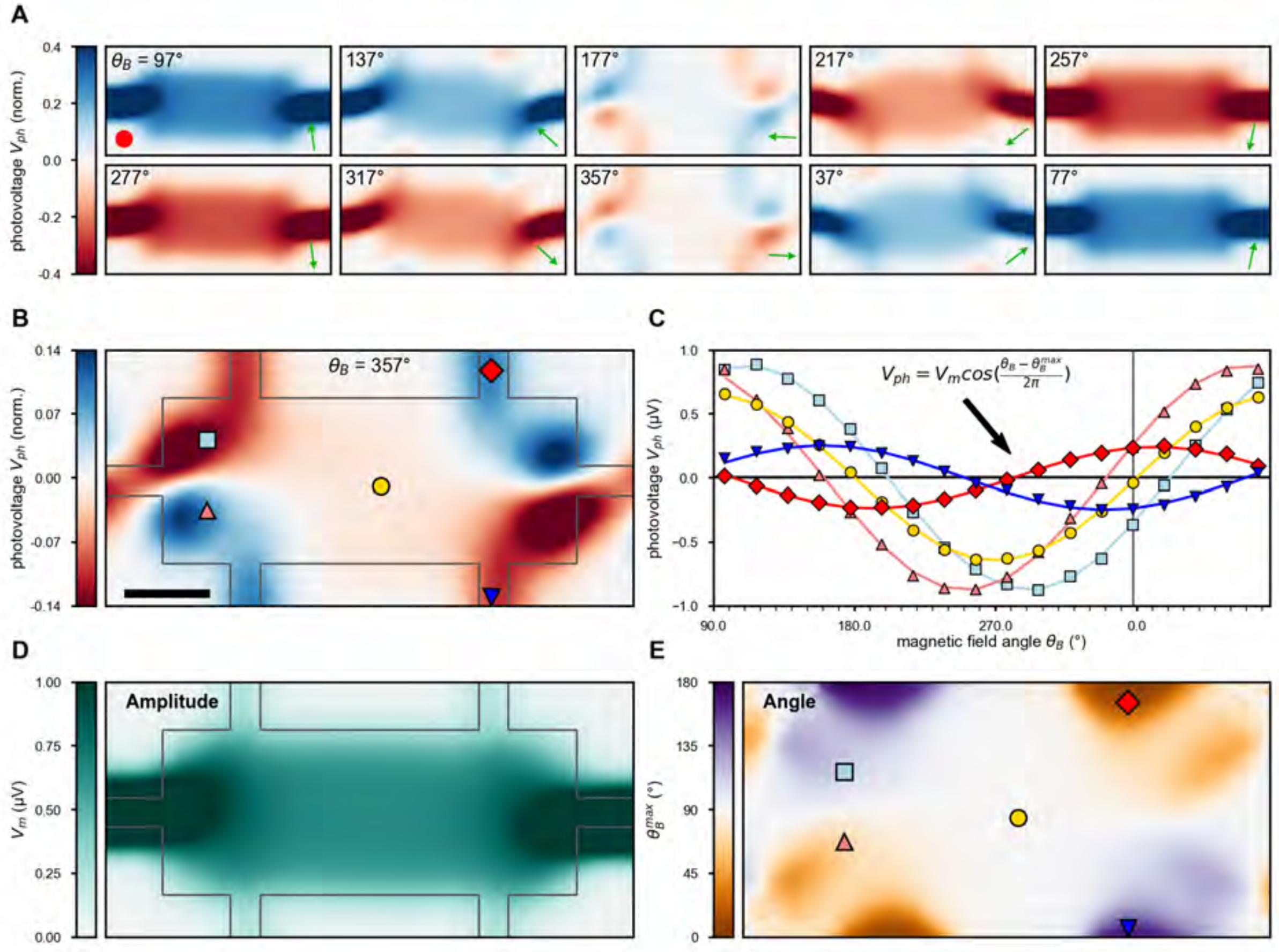

**Fig. 2.** *Data-intensive imaging of the magnetic field-dependent photovoltage in a micromagnetic heterostructure device.* ***A***, Magnetic field-dependent photovoltage maps at 10 different magnetic field angles. $B$-field direction, $\theta_B$, is labelled top left corner and is indicated by green arrows. Red circle indicates the FWHM of the beam spot (FWHM = 27 μm). ***B***, Detailed view of the anomalous photovoltage features at $\theta_B$ = 357°. Scale bar 50 microns for all images. ***C***, $V_{ph}$ vs. $\theta_B$ for 5 points marked in ***B***. Data points share the same colors and shapes as in ***B***. At each point in space, $V_{ph}$ vs. $\theta_B$ is fit to the function $V_{ph}(\theta_B) = V_m \cos(\theta_B - \theta_B^{max})$, shown as corresponding solid lines. ***D***, Image of the sinusoidal fit amplitude $V_m$ at all points in space. ***E***, Image of the angular phase shift of the fit $\theta_B^{max}$ relative to $\theta_B$ = 0° at all points in space. Marked points are the same as those in ***B***, corresponding to the $V_{ph}$ vs. $\theta_B$ data in ***C***.

The flowlines observed in Figure 3 meander smoothly within the Pt film conforming to the device boundaries. These flow lines indicate the direction of *local* (photo-induced) charge flow that maximizes the *global* photovoltage response measured at current-drawing contacts. As such, they naturally explain the photovoltage polarity changes observed at corners and near the additional side contacts in Figure 2B: local charge current that is aligned to the flow lines generates maximum positive photovoltage (red diamond Figure 2B), while anti-aligned local charge current generates negative photovoltage (blue triangle Figure 2B). Conversely, $V_{ph}$ becomes zero when the local photocurrent is perpendicular to the flowlines.

Reminiscent of fluid streamlines through a wind tunnel, these current flowlines do not terminate within the Pt film so that total flowline flux is conserved. As a result, the current streamlines are highly sensitive to the device geometry: flowlines in Figure 3B are uniform and dilute far from the boundaries but converge and bend into a high-density bunch where wide regions meet narrow constrictions. For a finite spot-size laser illumination, the density of current flowlines corresponds directly to the intensity of the photovoltage response. Indeed, as shown in Figure 3B, $V_{ph}$ is enhanced in regions of high current flowline density (e.g., along the narrow horizontal contacts) and suppressed in regions of low density.

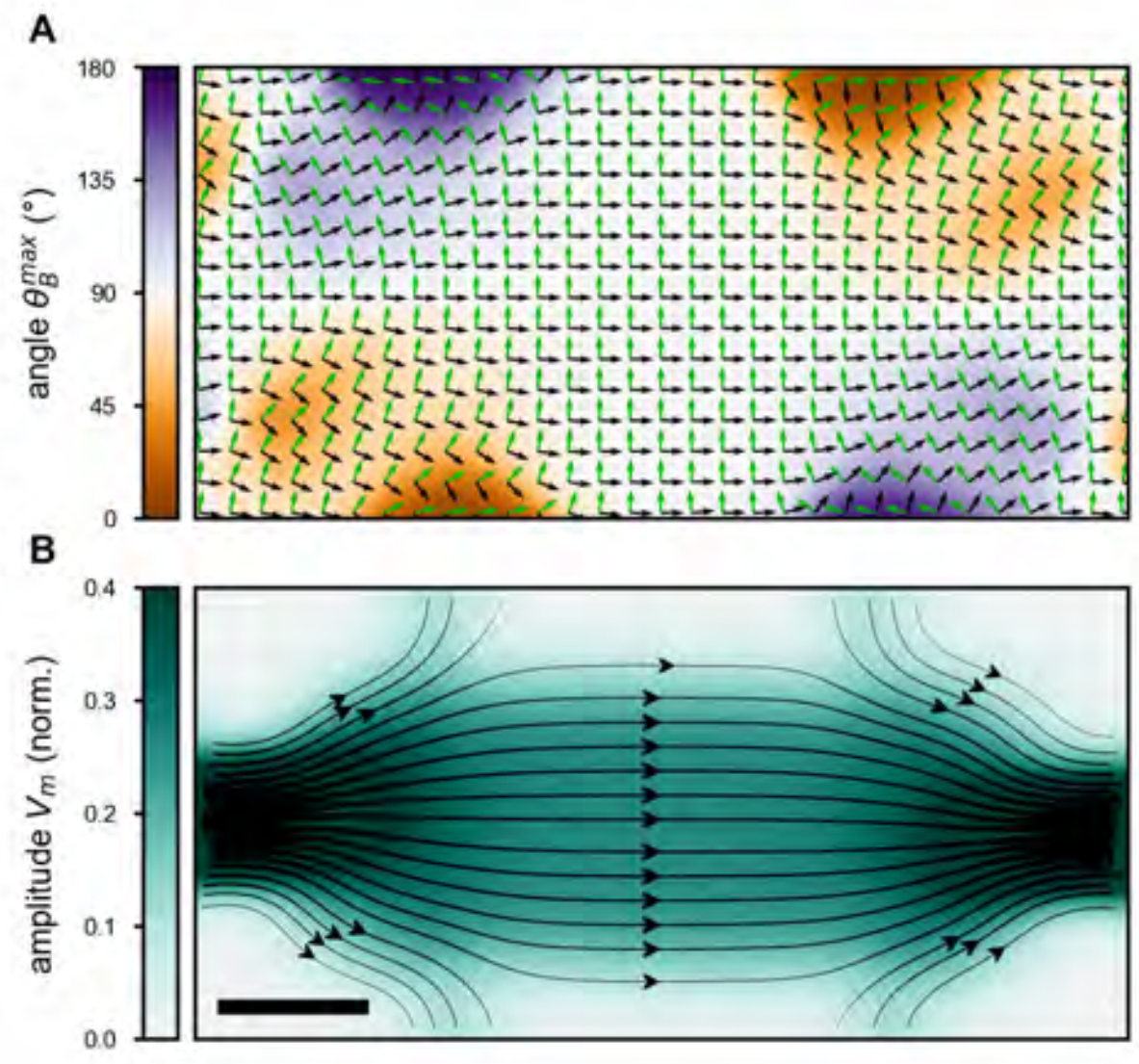

**Fig 3.** *Imaging the current flux through a micromagnetic heterostructure.* ***A***, Image of $\theta_B^{max}$ overlain with 2 vector fields: Green arrows indicate the direction of $\theta_B^{max}$ at all points in space. $\theta_B^{max}$ determines the direction of the magnetic field that yields maximum detectable photovoltage signal. Black arrows indicate local charge current density resulting from thermospintronic response. ***B***, Continuous flow-field interpolation of the current density vector field in ***A*** overlaying the sinusoidal fit amplitude $V_m$ at all points in space. Scale bar 50 microns.

Current flowlines (and the global photoresponse) can be manipulated far away from device boundaries in much the same way as fluid streamlines are guided to flow around an airplane wing. To test this, we fabricated cross-sectional wing shapes, or electrofoils, within the micromagnetic devices. With the exception of an un-patterned blank device (Figure 4A left), electrofoils were created by removing regions of the Pt film in the shape of aerodynamic Clark Y airfoils (Figure 4); The electrofoils exhibit a convex upper profile and flat lower surface (gray outlines Figure 4A), and each is fabricated with increasing angle of attack. Similar to the Hall bar device, we measure $V_{ph}$ vs. $\theta_B$ and fit the resulting sinusoidal characteristics to obtain $V_m(x,y)$ and $\theta_B^{max}(x,y)$ (Figure 4A,B). From these maps, we then produce images of the current flow through the electrofoil devices (Figure 4C) using the protocol described above.

Figure 4D and 4E show detailed images of the current flowing over the top surface of the $35^0$ electrofoil. At such high angle of attack, flowlines curve sharply to traverse the electrofoil (black lines Figure 4D), creating regions of high- and low-density current flow. The photovoltage amplitude $V_m$ in Figure 4D reaches a peak value above the leading (left) edge of the electrofoil. By introducing the asymmetric cambered boundary, flowlines are forced to curve more sharply around the leading edge of the electrofoil compared to the trailing edge. In this way, we gradually manipulate the flowline density - and thus photovoltage intensity - along the length of the electrofoil surface.

The flowlines obtained through our data-intensive imaging technique give a unique fingerprint for each device (and each electrofoil), clearly identifying the lines along which current naturally flows. We emphasize that the flowline patterns, while device specific, are independent of the mechanism through which local currents are generated. Instead, when a local current density $\vec{j}_c$ is induced (e.g., by local laser illumination), it acts as a local electromotive force that drives ambient charge carriers along the unique flow field determined by the device geometry. This flowline pattern is in full concordance with that predicted by the Shockley-Ramo theorem[22,23] for conductors[24], wherein a local electromotive force creates a global diffusion current that flows into the global contacts giving the measured photovoltage $V_{ph} \propto \int \vec{j}_c(\vec{r}) \cdot \vec{S}(\vec{r}) d^2\vec{r}$. Here $\vec{S}(\vec{r})$ is the smoothly varying current flowline pattern bounded by the edges of the device (in our case, the conducting Pt plane). The experimental data of Figure 4D directly visualizes these natural flowlines inside the electrofoil device.

Although the current streamline field $\vec{S}(\vec{r})$ is usually assumed to be a mathematical concept - akin to the very useful concept of gravitational or electrical potential - our data-intensive optoelectronic technique provides the first images of these natural flowlines inside an electronic device. Under laser illumination, directional control of *local* spin and charge current (through *B*-field manipulation) enables the detailed visualization of these *global* current streamline fields. Such flow fields are not unique to micromagnetic devices and can be used as a direct probe to image unusual charge flow in several recently proposed interacting electronic systems, such as graphene[25,26] and $PdCoO_2$[27]. As in aerodynamics, flowlines may be modified in the presence of viscosity, which manifests as a frictional force between fluid layers. While the global consequences of viscous electron liquid are being explored[4,8,28-30], the spatially resolved details of unconventional streamline fields $\vec{S}(\vec{r})$ have so far been out-of-reach. Current flux imaging is a robust new experimental tool for detailed visualization of unconventional flow. Yet, perhaps most compelling, is its application as an electronic 'wind tunnel' for the study of charge, spin, and heat flow in nanoscale electronic devices.

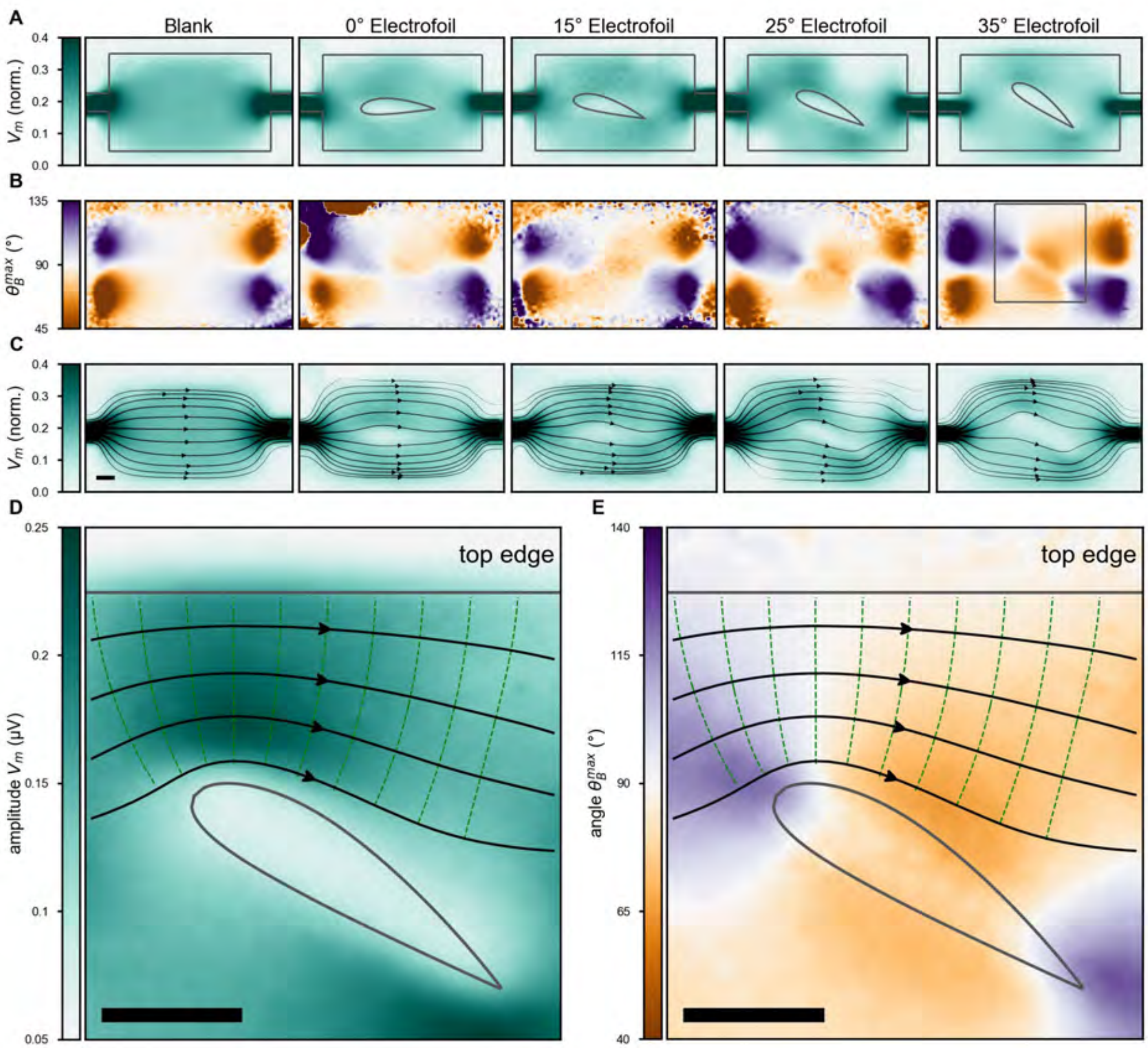

**Fig 4.** *Contorting, compressing and decompressing current streamlines through a micromagnetic electrofoil.* ***A***, Images of the sinusoidal fit amplitude $V_m$ at all points in space for 5 different magnetic heterostructure devices. The blank device is a two-terminal rectangular plane (500 µm x 250 µm). Electrofoil devices are two-terminal planes from which Pt is removed in the shape of aerodynamic airfoils with increasing angles of attack (labelled above), as outlined with solid lines. ***B***, Images of the angular phase shift of the fit $\theta_B^{max}$ relative to $\theta_B = 0^o$ at all points in space for each of the electrofoil devices. ***C***, Continuous flow-field interpolation of the current density vector field in ***B*** overlaying the sinusoidal fit amplitude $V_m$ at all points in space. Interpolated flow lines are weighted by the amplitude $V_m$. ***D*** and ***E***, Zoomed in view of $V_m(x, y)$ and $\theta_B^{max}(x, y)$ for the $35^0$ electrofoil device. Solid black lines indicate current flow. Scale bar 50 microns.

**Acknowledgments:**

The authors would like to acknowledge valuable discussions with Ilya Krivorotov and Flip Tanedo. D.M., M.G., and N.M.G. were supported by the National Science Foundation Division of Materials Research CAREER award no. 1651247. N.M.G. acknowledges support through a Cottrell Scholar Award, and through the Canadian Institute for Advanced Research (CIFAR) Azrieli Global Scholar Award. D.M., M.G., M.L., M.A., J.L., and J.S., were supported as part of the SHINES, an Energy Frontier Research Center funded by the U.S. Department of Energy, Office of Science, Basic Energy Sciences under Award No. SC0012670. J.C.W.S. acknowledges support from Singapore National Research Foundation (NRF) under NRF fellowship award number NRF-NRFF2016-05, and Singapore MOE Academic Research Fund Tier 3 Grant MOE2018-T3-1-002. All authors contributed to the writing of the manuscript. N.M.G. conceived the experiment, as well as supervised the analysis and interpretation with additional input from V.A., J.S., and J.C.W.S. D.M. and M.G. executed detailed optoelectronic measurements on devices and materials made in collaboration with M.L., M.A., J.L., and J.S. All authors evaluated the data analysis and content of the manuscript. Authors declare no competing interests. All raw data and material analysis code (python) are available upon reasonable request and material production is outlined in the supplementary text.

## SUPPLEMENT:

### S1. Device Fabrication and Materials Characterization

The devices used in this work were designed using highly characterized micromagnetic heterostructures. As described below, detailed magnetic characterization of these metal/magnetic insulator devices has previously been reported in several manuscripts by our group and others. Our samples consist of thin platinum films (5 nm) patterned on a thin film of yttrium iron garnet (YIG) on top of gallium gadolinium garnet (GGG). The 80 nm YIG films are grown via pulsed laser deposition on GGG substrates with (110) crystal orientation. The details of the growth for the YIG thin films was described in an earlier report[31]. After confirming the atomic flatness and room temperature magnetic properties of the films we used e-beam lithography with PMMA resist topped with Elektra 92 (SX AR-PC 5000/90.2) conductive resist to create the different device patterns on the thin film surface. The pattern used for the electrofoil was designed using designCAD 2000 to mimic the Clark Y airfoil design. Each was rotated at different angles relative to the $0^o$ (horizontal) design. After developing the patterns, we deposited the film in a sputtering chamber with base pressure of $5 \times 10^{-8}$ Torr for deposition of 5nm Pt. This was then followed by liftoff of the negative exposure features in acetone.

The coercive magnetic field of thin YIG samples has been found to be very small in both the in-plane[32] and out-of-plane[33] directions, measuring less than 0.2 T in all cases for films as thin as 4 nm and as thick as 35 nm[32]. External magnetic fields that exceed the coercive field saturate the magnetization of the YIG layer. The uniform $B$ = 520 mT magnetic field in our measurement was chosen to sufficiently overcome the coercive field, thus ensuring that the magnetization - and therefore the spin polarization vector of any out-of-plane spin currents in the system - is aligned to the external *B*-field.

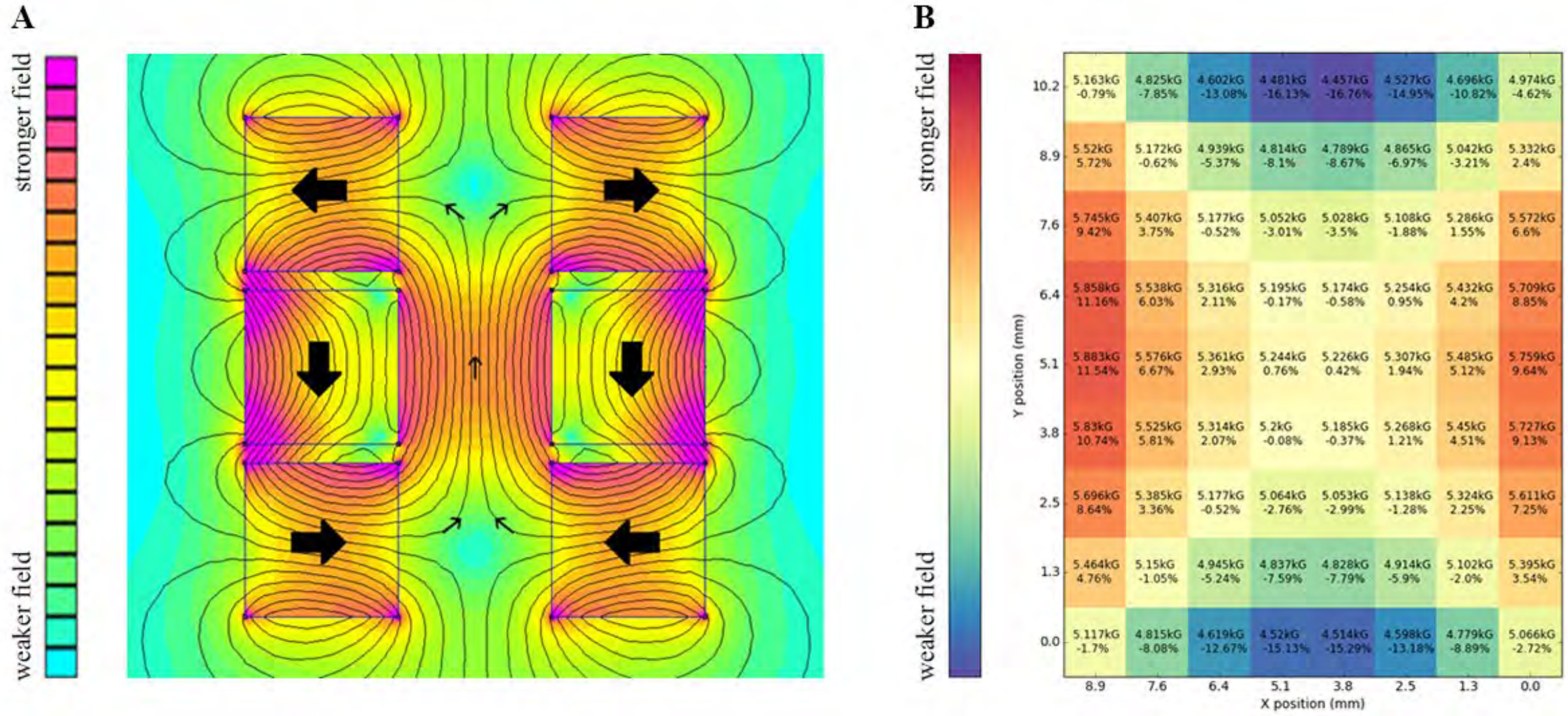

**Fig. S1.** *Magnetic field strength vs. spatial position.* **A**, Field simulation of the Halbach array and the resulting magnetic field as illustrated by the magnetic field lines shown. Solid square lines represent outline of N52 neodymium magnets. Arrows indicate direction of the neodymium magnet's field. Curves are simulations generated by Finite Element Method Magnetics software. **B**, Readings of the magnetic field strength as a function of distance in the two-dimensional sample plane. In the center area, at the sample position, the field varies by less than 1%, while there is less than 20 % variance throughout the entire square-centimeter area pictured above.

### S2. Data-Intensive Imaging using Scanning Magneto-photovoltage Microscopy

All measurements were performed using scanning magneto-photovoltage microscopy (SMPM) and analyzed using scalable methods of multi-dimensional data analysis developed in our previous work[34,35]. We introduce SMPM here for the first time and, in Section S2.1, we present a detailed description of its setup and application. Special attention is paid to the use of a Halbach array to produce a highly uniform and fully rotating magnetic field, an optics system that focuses a laser beam onto the device, and details of the automated movement and rotation of the magnetic field. Section S2.2 presents details on the data-intensive acquisition and analysis of the SMPM imaging data.

#### S2.1. Scanning Magneto-Photovoltage Microscopy

SMPM is a new experimental tool designed to aid in the collection of large and complex data sets resulting from measurements of the photoresponse in novel micromagnetic materials. In this work, a scanning laser is used to generate a temperature gradient and induce the Longitudinal Spin Seebeck Effect (LSSE, discussed in detail in section S3) in a micromagnetic device while a broad parameter space is explored. For example, SMPM is automated to adjust power (via a neutral density filter), the 2D beam spot position (via a galvanometer), beam polarization, in-plane (x-y plane) magnetic field angle, and the out-of-plane magnetic field angle (x-z plane). SMPM is programmed to repeatedly scan over a given device while exploring various experimental parameters, resulting in large, discrete datasets called *run*s. Each run contains the back-reflection and photovoltage-response amplitudes for measurements taken at each configuration of the given experimental parameters (e.g., magnetic field direction). High resolution measurements of all available parameter space are unfeasible, and a typical run is reduced to studying the interdependence of 3 or 4 parameters. In this work, we focus on spatial images and *in-plane* magnetic field orientation.

SMPM utilizes a high frequency mode-locked pulse laser to generate a photovoltage response, a rotatable Halbach array to produce a magnetic field, a lock-in amplifier, and home-built DAQ setup[34] to record the photovoltage response. The 830 nm wavelength, mode-locked, titanium-sapphire pulsed laser, with 200 fs pulses and a 76 MHz repetition rate is used to produce a spatially localized thermal gradient, which gives rise to a directional net spin-current density due to the LSSE. A set of two galvo-controlled mirrors are used to scan the beam while a set of optics focus the beam on the back of the objective, resulting in a focused, scanning beam-spot. The focused Gaussian beam spot was measured using a knife-edge technique and found to have a full-width-half-max (FWHM) of 23 microns at the device. This large beam spot is used to integrate the extremely weak LSSE signal in the *linear response regime*. In other words, we confine our measurements to parameters in which photovoltage increases linearly with power, and for the measurements shown in the main text, average beam power is held at or below 30 mW. In addition to recording the photoresponse, a back-reflection amplitude is recorded for each measurement by introducing a 50/50 beam-splitter and photodiode to the beam path ahead of the galvo-controlled mirrors. A lock-in amplifier measures the relative voltage between two terminals and is synchronized to an optical chopper at low frequency.

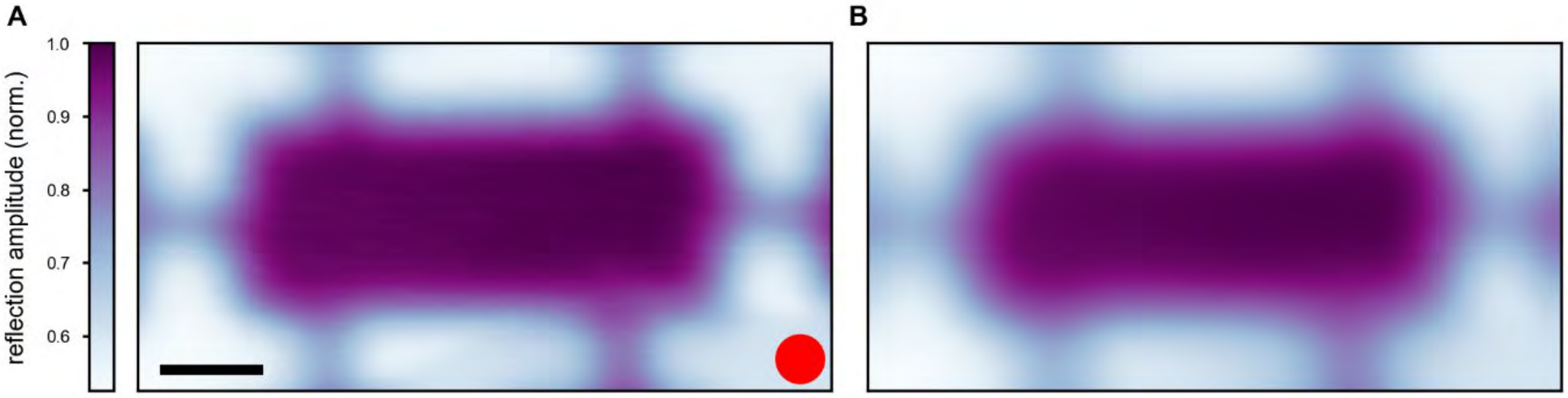

**Fig. S2.** *Reflection images of the Hall bar device before and after drift-correction.* ***A***, Back-reflection image of the Hall bar device prior to gaussian filtering and drift correction. ***B***, Back-reflection image after drift correction. Scale bar is 50 microns long while the beam spot size is 27 microns FWHM. Small horizontal lines in ***A*** but not ***B*** are the result of 4Hz optical chopper interference.

Micromagnetic devices are mounted to a small PCB mount within the highly uniform B-field area of the Halbach array. The array, made from 6 N52 neodymium magnets precisely arranged to control the magnetic field (see figure S1A), measures B = 0.52 T at the sample position (see figure S1B). To allow for 90º out-of-plane rotation and 360º in-plane rotation, the array is attached to a rotation mount (Thor Labs stepper motor rotation mount), which in turn is mounted to a rotation stage (Thor Labs heavy duty rotation stage with stepper motor). As shown in figure S1B, measurements of the magnetic field within the active region of the Halbach array show remarkable uniformity over an area of about 9 $cm^2$ in the center of the array.

For all devices, extensive orientation tests were carried out to ensure correct analysis of the multi-dimensional data. During this process, the device orientation was unambiguously established and magnetic field angles relative to the device were determined. This crucial process allows verification of the signs of the fields and currents in the system and provides a precise definition of $\theta_B^{max}$ used throughout this paper.

#### S2.2. Data-Intensive Imaging and Data Analysis

SMPM images are collected by a home-built data acquisition (DAQ) system and are stored and analyzed automatically using a series of python scripts, all introduced in (34). All automation tools are home-written using Python and are available upon request. As only the spatial location of the beam and in-plane magnetic field angle are used as parameters for the work presented here, the 3-dimensional data sets are easily visualized as a set of 2D images taken under different magnetic field orientations. The reflection amplitudes are stored in a separate datacube and are used to calibrate the SMPM and account for drift over long periods.

A typical run, which acquires ~$10^5$ photovoltage and reflection measurements, takes between 5 and 15 days to complete, depending on the data resolution. To account for possible sample drifting in over this time, drift correction is performed in the same manner as described in[34], the results for which can be seen in figure S2. Note that during this process a 2D gaussian filter is run on each slide of the datacubes, which integrates out small features that can be seen in Figure S2A but not in figure S2B.

All imaging data and analysis scripts used here are contained in a single Jupyter notebook, which is available upon request. The scripts contain well documented python code used for drift correction, Seebeck Effect extraction, and figure production. The analysis software relies heavily on the Matplotlib, NumPy, SciPy, and FiPy libraries.

**S3. Thermoelectric and Thermospintronic Response in Micromagnetic Pt/YIG Devices**

While the magnetic field-dependent photoresponse in Pt/YIG involves several effects, prior work[19,36-40] has demonstrated that, given the large out-of-plane temperature gradient, one effect dominates: The Longitudinal Spin Seebeck Effect (LSSE). While the conventional Seebeck (thermoelectric) effect can be measured directly, the LSSE can only be detected via the Inverse Spin Hall Effect (ISHE), all described below.

In Section S4 below, we show that our data-intensive technique (SMPM) gives a unique window into the intrinsic photoresponse: By separating out the magnetic field-*dependent* response from the magnetic field-*independent* response, we are able to isolate conventional thermoelectric response from thermospintronic response. The magnetic field-*independent* Seebeck effect is concerned only with the temperature gradient between the aluminum contacts on either side of the device and will be discussed in section S3.1. The magnetic field-*dependent* Longitudinal Spin-Seebeck Effect produces an out-of-plane spin-current that flows anti-parallel to the temperature gradient across the Pt/YIG film and is discussed in section S3.2. The process in which a spin-current propagating in a device with a net magnetization generates an electric current transverse to both is called the Inverse Spin Hall Effect and is also discussed in section S3.2.

S3.1. Conventional Thermoelectric (Seebeck) Effect

The conventional Seebeck effect, discovered by Alessandro Volta in 1794[41] while independently rediscovered by and named after Johann Seebeck in 1821, refers to the generation of a voltage across a conductor subject to a parallel temperature gradient, $V_{Se} = -S\vec{\nabla}T$. The underlying physical process driving this effect stems from the progressively higher kinetic energy imparted to charge carriers in hotter regions of a conductor, leading to a gradient in the mean free path of charge carriers and an overall outward flow. If there are metal junctions on either side of the heat source interrupting this electron flow, the difference between the Seebeck coefficients for the two metals, encoding the temperature dependence of each material's chemical potential, will cause a disproportionate charge build-up at the junction furthest from the heat source. While the Seebeck effect is described as an electric field, $E = -S * \nabla T$, across the device, it is important to note that the measurable voltage is the result of an electromotive force which seeks to restore equilibrium after the Seebeck effect disturbs local charge densities. This effect occurs independent of any magnetization or external magnetic field.

**S3.2. Thermospintronic Response: The Longitudinal Spin Seebeck Effect (LSSE)**

The Longitudinal Spin Seebeck Effect (LSSE) is the generation of a spin potential and an accompanying spin current in a magnetic material by means of an out-of-plane temperature gradient[10,19,40,42-43]. A conventional measurement of the LSSE requires one or more Peltier modules to control the temperature at the top and bottom of the device. This allows simple calculation of the temperature gradient across the magnetic material and will produce LSSE response without the Seebeck effect. Here we attempt to generalize these experiments by accounting for spatial variation in the effects.

While the mathematical form for $j_S$ can be quite complicated, taking into account the density of spin carriers as well as the effective spin-mixing conductance at the interface between the magnetic and metallic layers, crystal symmetry dictates it is spatially independent, so it is expected that a consistent temperature gradient will produce the same spin current at any given point on a device[42]. It is found that for Pt/YIG, the spin current will flow anti-parallel to the temperature gradient[39,44]. As a laser is used as a source for the temperature gradient, heating the platinum layer first, spin currents should experience a net flow from the platinum into the YIG.

In the classical Hall effect, a current running through a conductor interacts with a transverse magnetic field to produce a potential difference across the conductor which is orthogonal to both. The spin-current counterpart to this effect, the Spin Hall Effect (SHE), is a process in which an incoming charge current produces a spin current transverse to it. The SHE is observed in quantum heterostructure devices in which a normal metal with sufficiently high spin-orbit coupling (SOC) is bonded to either a ferromagnet or ferrimagnetic insulator (FMI), such as YIG. As the charge carriers propagate through the normal metal, the SHE produces a small torque which drives carriers with opposite spins in opposite directions[15]. The Inverse Spin Hall effect (ISHE) is the reverse process, in which a spin current traveling normal to the interface between a metal and FMI is polarized transverse to the spin current, resulting in a charge current orthogonal to both[15,17,39]. As a spin-current propagates out of plane it deflects charge carriers in the metal to conserve momentum. As charge-carriers of any spin are deflected in the same direction, depending only on magnetization and spin-current directions, an electric current is generated across the metal.

As platinum has strong SOC, the Inverse Spin-Hall effect (ISHE) will convert a spin current traveling normal to its surface into a charge current orthogonal to both the spin current and the spin polarization vector. This is, in general, described mathematically by $j_C = D_{ISHE}(j_S \times \sigma)$, where σ is the spin polarization vector of the material, $j_S$ is the spin current from the LSSE described above, and $D_{ISHE} = \theta_{SH}\left(\frac{h}{4\pi e}\right)$ is a material dependent constant. For a 15 nm platinum device at room temperature, $\theta_{SH}$ is very small and $D_{ISHE} \approx 1 \times 10^{-17}$Wb (45). For this experiment, the spin current points along the negative z axis (from Pt to YIG) and the magnetic field is kept purely in-plane. The generated charge current will also be in-plane, with angle $\theta_B - \frac{\pi}{2}$. Thus, in the center of a large, symmetric device, it is expected that a maximum photoresponse would be measured when the external magnetic field angle is rotated to the angle $\theta_B^{max} = \frac{\pi}{2}$ where a maximum potential is measured when the current propagates in the positive *x*-direction.

**S4. Experimental Procedure**

In this section, the experimental procedure is discussed to include a careful walk-through of the data acquisition, data analysis and image presentation of the main manuscript. Section S4.1 describes the use of SMPM, data-cubes, and the analysis toolbox modified from those developed in (34). With section 4.1 providing details on the setup of the experiments, sections S4.2 through S4.4 describe the steps taken from raw imaging data to spatial maps of the current flowlines. Section S4.2 discusses the handling of raw, drift-corrected data sets, including both reflection and photoresponse data-cubes. Section S4.3 describes the process taken to extract the Seebeck Effect from the photoresponse data-cube and subsequently separate the (magnetic field-*independent*) Seebeck and (magnetic field-*dependent*) Spin-Seebeck Effects into separate data-cubes to be analyzed individually. This section also contains results from Seebeck measurements, which were omitted from the main text. Finally, section S4.4 discusses the steps in the analysis of the Spin-Seebeck (field-dependent) data-cubes, while the reader is encouraged to read section S5 for fit accuracy details and elimination of additional effects.

**S4.1. Data Acquisition**

For the Hall bar Pt/YIG, terraced Hall-cross (shown below), and un-patterned devices, the FWHM beam spot was measured to be 23 microns. As the power was kept at 30 mW for all measurements, these devices were measured with the laser power density at $1.0 \times 10^4$ W/cm$^2$. The beam spot for the electrofoil devices was larger, with a FWHM at 54 microns, yielding a power density of $2.3 \times 10^3$ W/cm$^2$. The magnetic field was held purely in-plane for the duration of all measurements in this project.

**S4.2. Raw Data and Preliminary Analysis**

Once a measurement has been completed, and drift correction and other pre-analysis data-manipulation techniques have been performed as described in sections S2 and S4.1, data analysis begins with analyzing the photovoltage datacubes. The raw datacubes are easily visualized as a series of 2D spatial maps of either the reflection or photoresponse amplitude resulting from illuminating a certain region of the device with the laser, given an evolving magnetic field orientation. The reflection and photovoltage data-cubes generated by our in-house DAQ setup during measurement are imported into a Jupyter notebook for python-assisted analysis. As the magnetic field does not affect the reflection amplitude, the reflection data-cubes are averaged across the magnetic-field axis to produce a single image of the platinum layer of the devices. Figure S3 shows the reflection amplitude for the Hall bar (figure S3A) and electrofoil devices introduced in the main text (figures S3 D-H) as well as 2 additional geometries studied here: an un-patterned Pt/YIG/GGG device (figure S3C) and a terraced Hall cross (figure S3B). The reflection images are used to guide the analysis of the photoresponse maps, providing an illustration of spatial displacement.

The first step in analyzing photovoltage datacubes is to display them in image form. Figure S4 is broken into four subfigures, each made of a series of raw photoresponse maps collected from various device geometries and varying by magnetic field-angle. The effects vary greatly depending upon the geometry, but when the field is parallel to the terminal axis, there is a consistently small photoresponse in the bulk of the device, while an orthogonal field produces a maximum (or minimum) response.

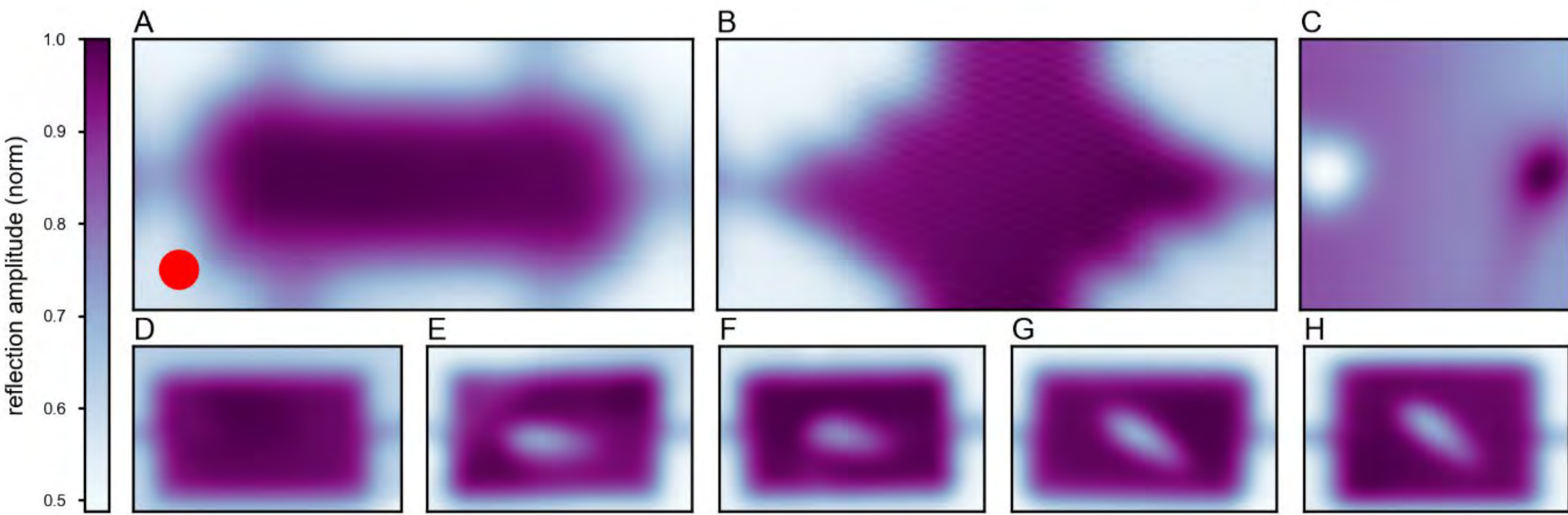

**Fig. S3.** *Back-reflection Amplitudes for various device geometries*. Images shown are displayed after drift-correction (see Section S2 and figure S2). All plots are normalized reflection amplitude maps, illustrating the geometry of a device's platinum layer, and are used for calibration and drift correction. ***A***, Hall bar device used throughout the main text. The beam spot is 27 microns FWHM. ***B***, Terraced Hall cross and ***C***, all platinum device, are introduced and referenced throughout this supplement. The remaining electrofoil devices are used throughout the text and supplement. ***D***, Large 2-contact device. ***E-H***, Electrofoil devices arranged by increasing angle of attack: 0º, 15º, 25º, and 35º.

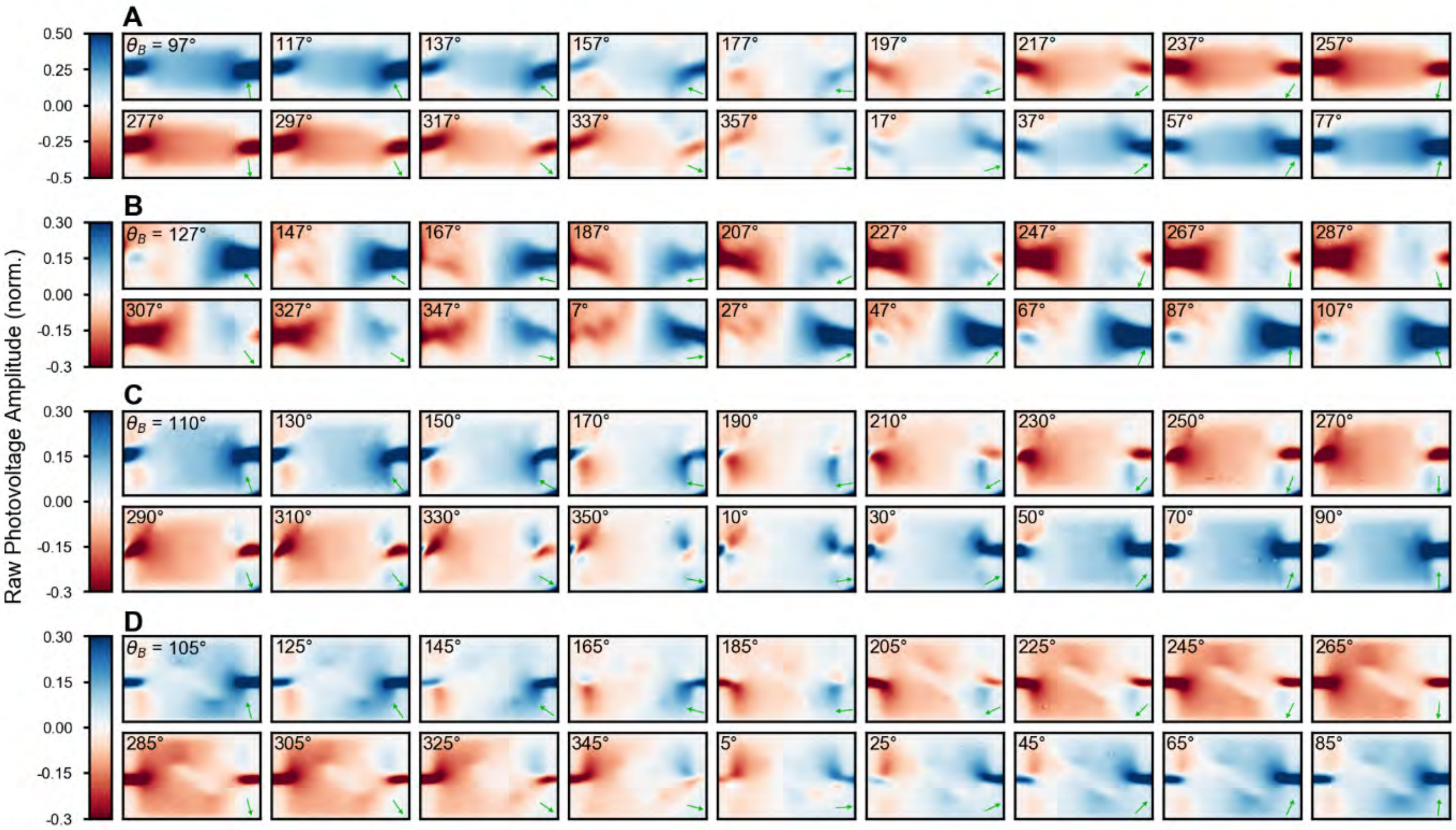

**Fig. S4.** *Raw photovoltage data-cubes.* Each subfigure contains 18 images, taken every 20º and represent the raw SMPM results. Four device geometries are depicted: ***A***, the Hall bar device used throughout the main text, ***B***, a terraced Hall cross, ***C***, the blank electrofoil device introduced in the main text, and ***D***, the 35º electrofoil device introduced in the main text. Magnetic field angle relative to terminal axis are indicated by $\theta_B$ in upper left corner of each image as well as green arrows in the bottom right corner. All images are normalized and saturated to highlight effects, as contact terminals generate a much larger signal than the bulk of the devices.

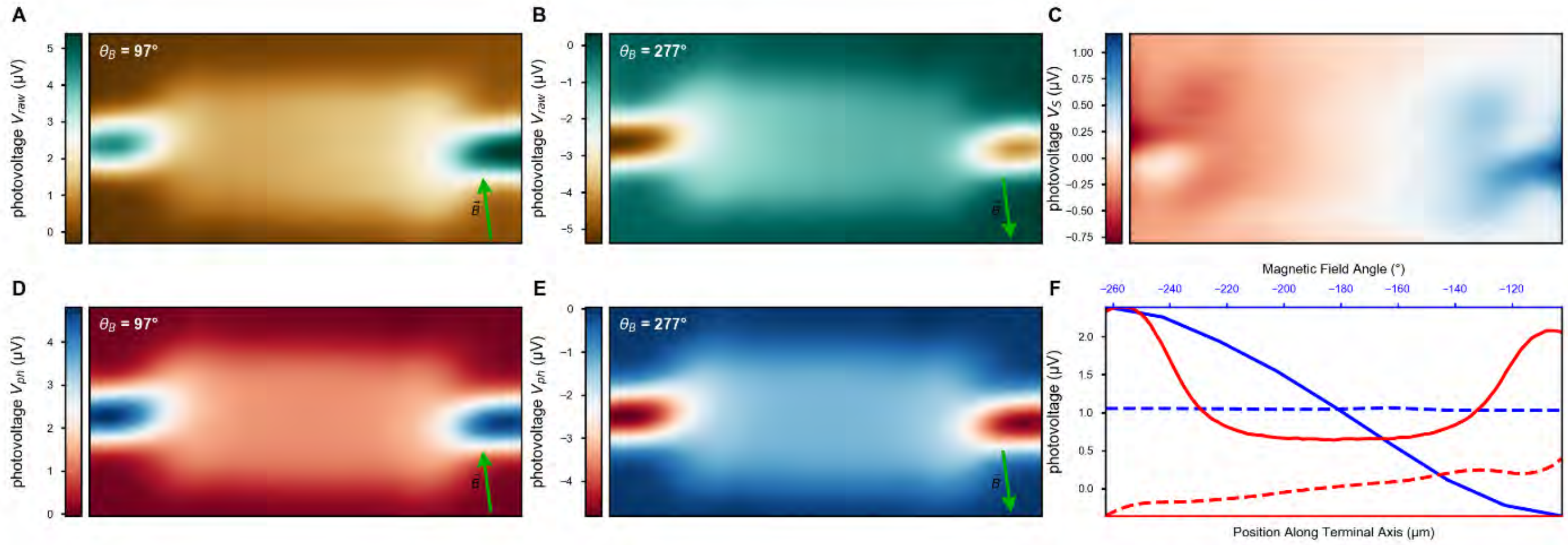

**Fig. S5.** *Extracting the classic thermoelectric (Seebeck) Effect.* Example illustration of the Seebeck extraction process using the same Hall bar sample as used throughout this work. Field orientation labeled by $\theta_B$ numbers in the upper left-hand corner and the directions of the green arrows. ***A***, Raw photovoltage amplitude map for the Hall-Bar device with a magnetic field orientation of 97°. ***B***, Raw photovoltage amplitude map for the Hall-Bar device with a magnetic field orientation of 277°. ***C***, A simple point-by-point average of the photovoltage amplitudes from 97° and 277° (the data in ***A*** & ***B***). ***D*** & ***E***, Subtracting the average (the data in ***c***) from the raw photovoltage data (the data in ***A*** & ***B***) produces $V_{ph}$. ***F***, Dashed lines represent a line trace of the Seebeck data while solid lines represent the spin-Seebeck data. The blue dashed line indicates the Seebeck effect is independent of the magnetic field.

**S4.3. Isolating the Thermoelectric (Seebeck) Effect**

As the platinum devices are connected to aluminum wires, a Seebeck voltage is expected to be measured along with the desired ISHE response, proportional to the temperature differences between the contacts. As the signal is expected to contain both magnetic field independent and dependent contributions, analysis continues with separating these effects. Figure S4 suggests that the field-dependent effects are sinusoidal in nature, thus the signal can be separated by taking the sum and difference of signals under opposite field orientation. This process is illustrated by figure S5, where the data in figures S5A and B are averaged together, point-by-point, to produce figure S5C, which in turn is subtracted from the data in figures S5A and B to produce figures S5D and E. As can be seen in the blue dashed line in figure S5F and discussed in Section S5, there is little variance in the Seebeck effect extracted data between different field orientations. These results are then subtracted from the raw data sets to produce the sinusoidally field-dependent data sets to be used later.

The Seebeck voltage is dependent only on the temperature gradient between the two aluminum wires on either side of the platinum device, meaning this signal will go from positive to negative as the laser scans from one terminal to the other. As can be seen by comparing the various devices in figure S6, the effect is sensitive to anisotropy in each device but has consistent behavior overall. The data is otherwise unremarkable and is consistent with the expectations of conventional thermoelectric response.

**S4.4. Analyzing Magnetic Field-Dependent Effects**

Once the field-independent data ($V_{\mathrm{sb}}$) is extracted, it is subtracted from the raw data set to produce the field-dependent data set ($V_{\mathrm{ph}}$). Figure S7 shows $V_{\mathrm{ph}}$ for different magnetic field angles for several different devices. Note the effect of the grounded contacts in the Hall bar device. The figure indicates that even when there are only probing contacts, there are still unusual edge effects when the magnetic field is parallel to the terminal axis. To examine this effect more thoroughly, each spatial point is fit to standard sinusoidal fitting function. After analyzing the results, the equation $V_{ph} = V_m \cdot cos(\theta_B - \theta_B^{max})$ is found to optimally fit the data, *with all residuals to the fit lying well below the noise floor of our measurements*. (section S5) Here $\mathrm{V_m}$ is the maximum amplitude of $V_{ph}$ and the cosine term captures the field-angle dependency. $\theta_B^{max}$, for a given spatial position, is found to be the angle the magnetic field takes, referenced from the positive x-axis, which results in a maximum measurable electric field pointing in the negative x-direction. Figures S8, S9, sand S10, replicas of figure 2 in the main text,

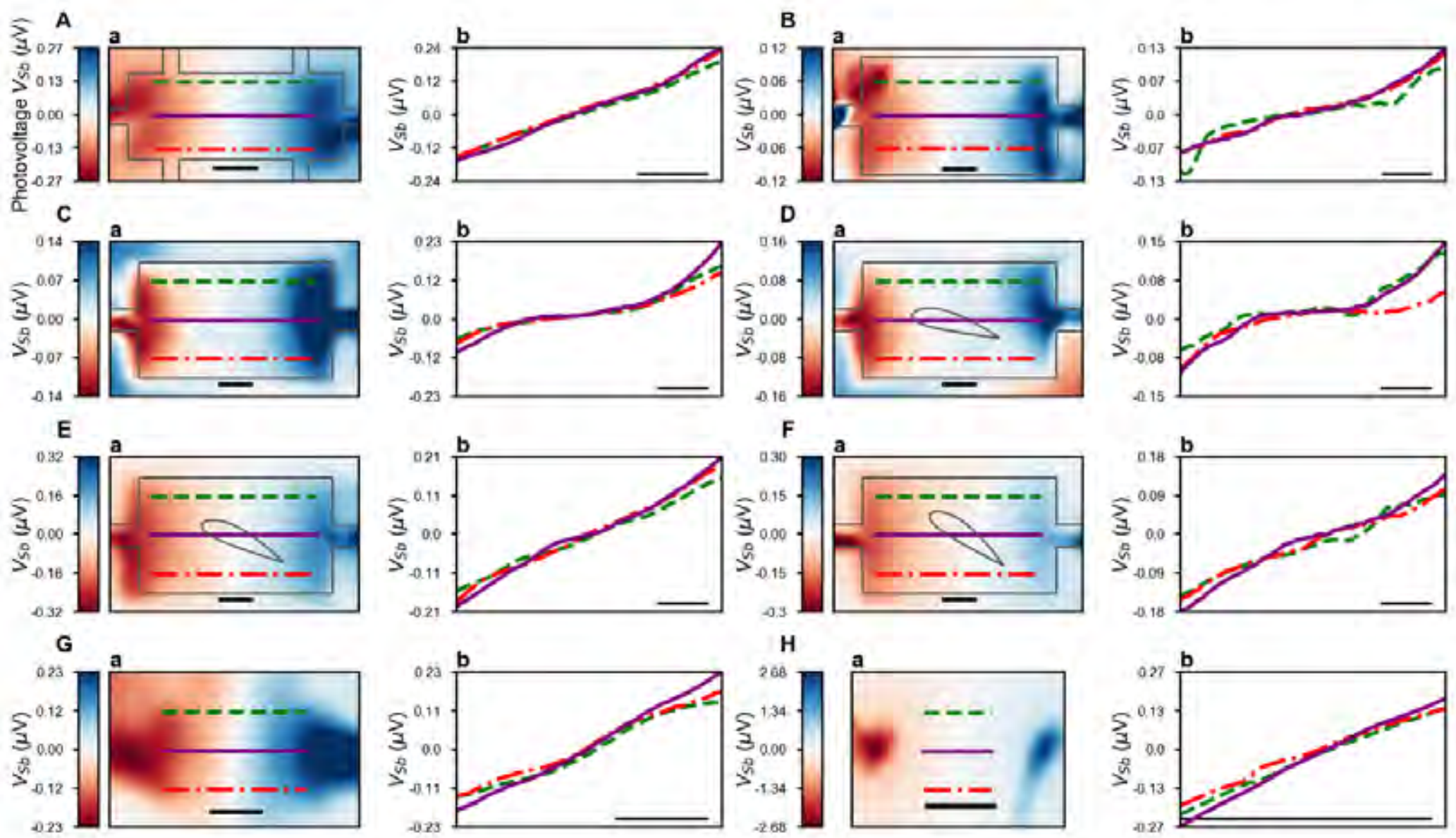

**Fig. S6.** *The conventional thermoelectric (Seebeck) effect for various geometries.* ***A-H***, Each panel contains a photovoltage map and line traces of the extracted Seebeck effect for a Pt/YIG device. ***A***, The Hall bar device used throughout the main text. ***B-F***, The electrofoil devices introduced in the main text and examined in figure 4 of the main text. ***B***, The blank electrofoil device, ***C***, the 0° electrofoil device, ***D***, the 15° electrofoil device, ***E***, the 25° electrofoil device, and ***F***, the 35° electrofoil device. ***G***, The terraced Hall cross and ***H***, the all-platinum device are used only in the supplement. ***a***, Photovoltage maps of the extracted Seebeck effect in various geometries. Solid black scale bars are 50 microns wide. Line traces are taken at three different positions along the devices and are represented by the colored solid and dashed used in both plots. ***b***, Spatial line traces of the photovoltage maps, depicting the consistency of the Seebeck effect across the devices.

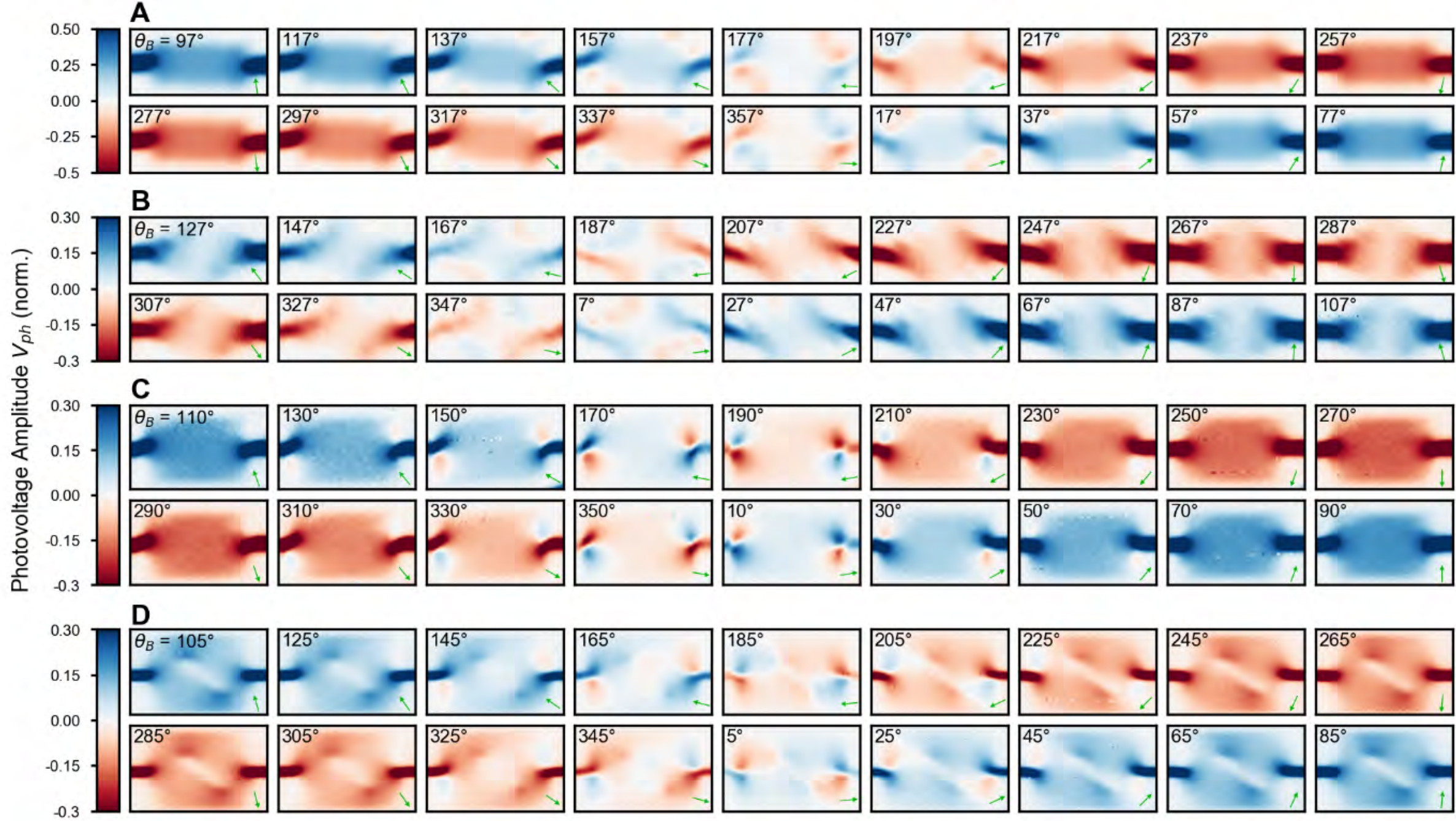

**Fig. S7.**

*Magnetic Field-dependent data-cubes.* Contains the same devices and angles as used in figure S4, now with the Seebeck effect extracted from each plot, as illustrated by figure S5. Each subplot, differentiated by device geometry, contains 18 photovoltage images, taken every 20°. ***A***, the Hall bar device used throughout the main text. ***B***, The terraced Hall cross device. ***C***, The blank electrofoil device, and ***D***, the 35° electrofoil device. All images are normalized and saturated to highlight their effects. Magnetic field angle relative to terminal axis is indicated by $\theta_B$ in upper left corner of each image and by green arrows in the bottom right corner.

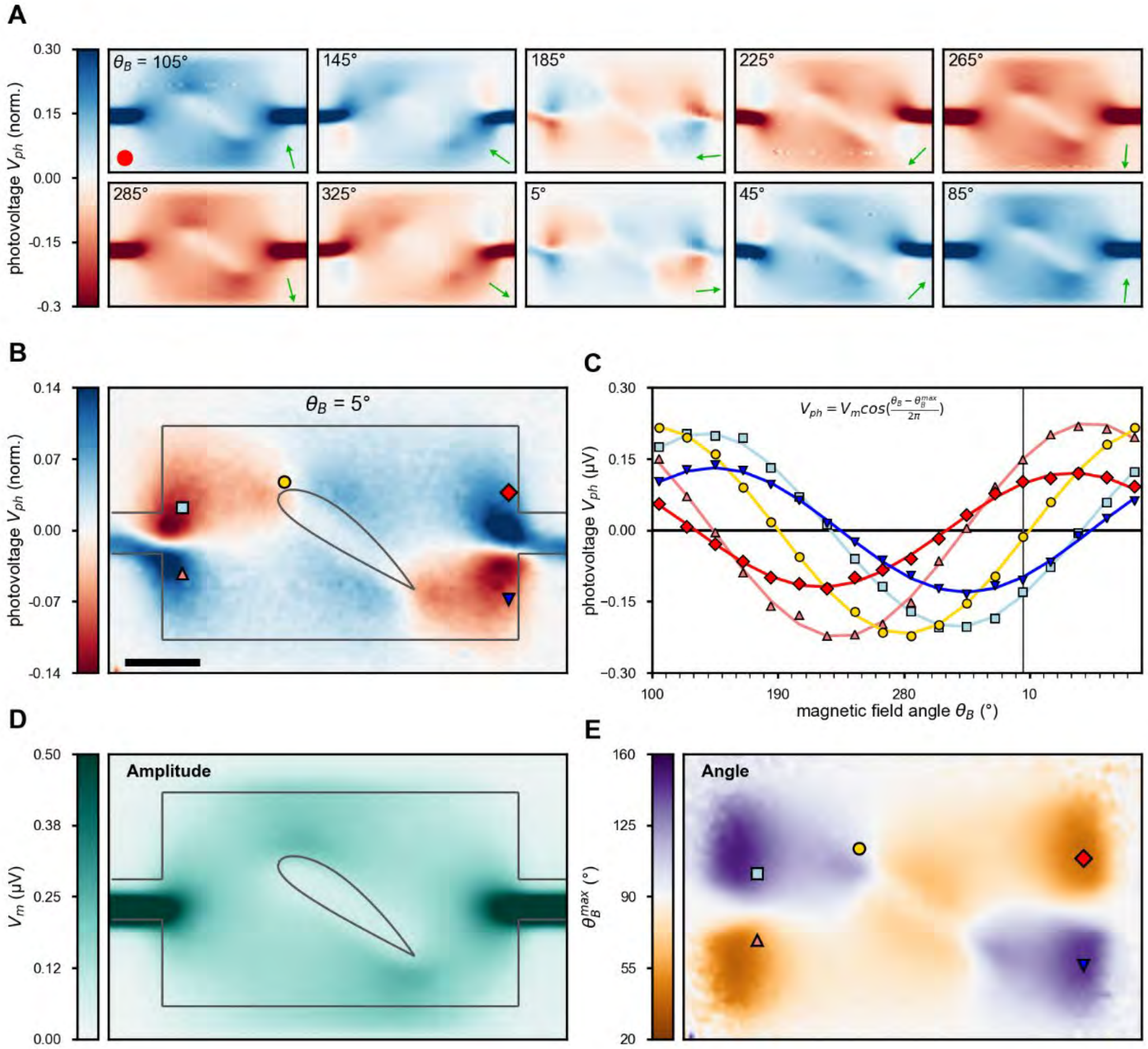

**Fig. S8.**

*Analyzing Field Dependent Effects in the 35º electrofoil device.* Replica of Figure 2 from the main text, using the 35º electrofoil device in place of the Hall bar device originally shown. ***A***, A subset of magnetic field-dependent photovoltage maps at 10 different magnetic field angles. $\theta_B$is labelled top left corner, *B*-field direction is indicated by green arrows. Red circle indicates the FWHM of the beam spot (FWHM = 54 μm). ***B***, Detailed view of the anomalous photovoltage features at $\theta_B$ = 357º. Scale bar 50 microns for all images. ***C***, $V_{ph}$ vs. $\theta_B$ for 5 points marked in ***B***. Data points share the same colors and shapes as in ***B***. At each point in space, $V_{ph}$ vs. $\theta_B$ is fit to the function $V_{ph}(\theta_B) = V_m \cos(\theta_B - \theta_B^{max})$, shown as corresponding solid lines. ***D***, Image of the sinusoidal fit amplitude $V_m$ at all points in space. ***E***, Image of the angular phase shift of the fit $\theta_B^{max}$ relative to $\theta_B$ = 0º at all points in space. Marked points are the same as those in ***B***, corresponding to the $V_{ph}$ vs. $\theta_B$ data in ***C***.

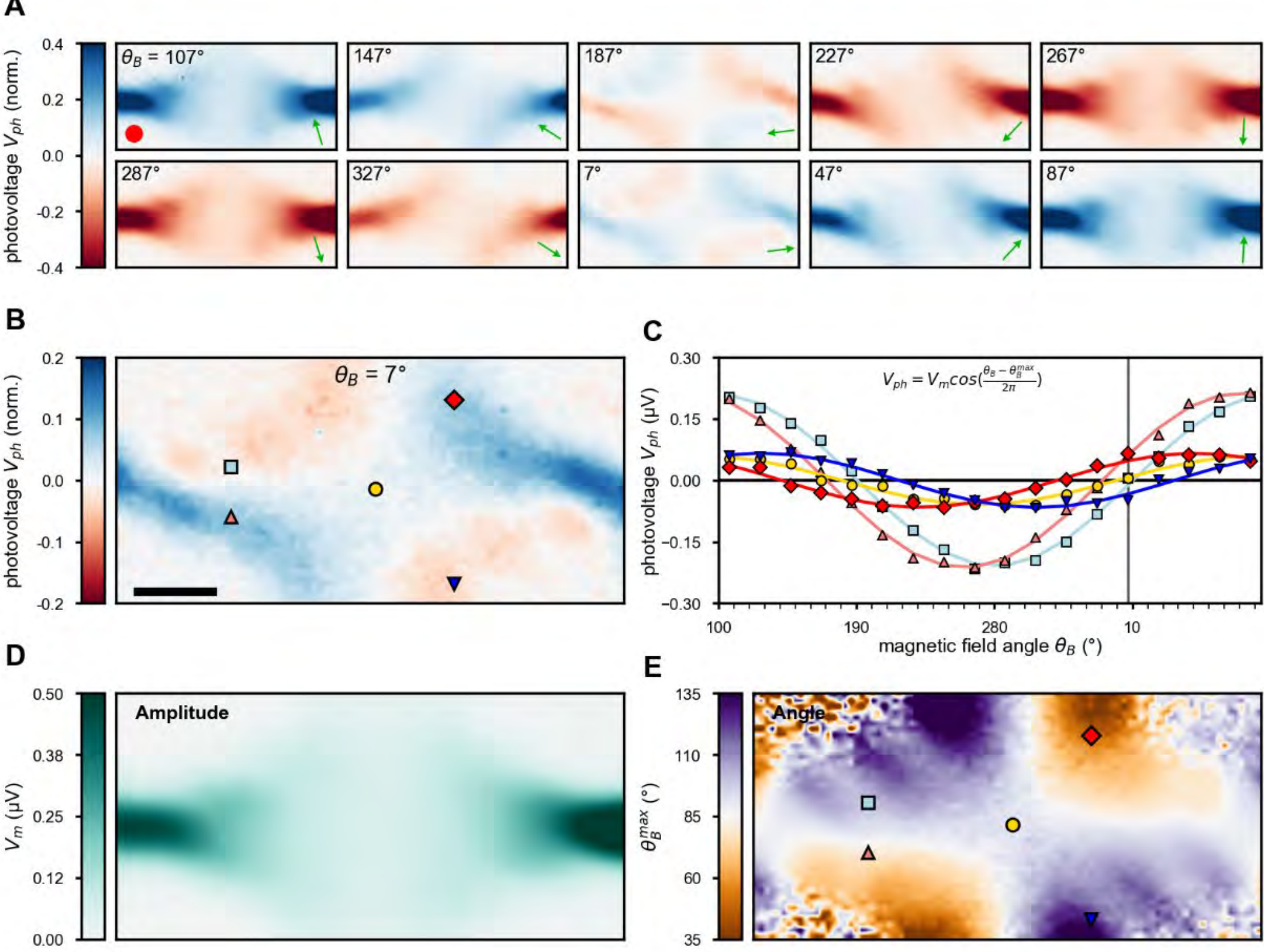

**Fig. S9.**

*Analyzing Field Dependent Effects in the Terraced Hall Cross device.* Replica of Figure 2 from the main text, using the terraced Hall cross Pt/YIG device in place of the Hall-bar device originally shown. ***A****,* A subset of magnetic field-dependent photovoltage maps at 10 different magnetic field angles. $\theta_B$is labelled top left corner, *B*-field direction is indicated by green arrows. Red circle indicates the FWHM of the beam spot (FWHM = 27 μm). ***B***, Detailed view of the anomalous photovoltage features at $\theta_B = 357^o$. Scale bar 50 microns for all images. ***C***, $V_{ph}$ vs. $\theta_B$ for 5 points marked in ***B***. Data points share the same colors and shapes as in ***B***. At each point in space, $V_{ph}$ vs. $\theta_B$ is fit to the function $V_{ph}(\theta_B) = V_m \cos(\theta_B - \theta_B^{max})$, shown as corresponding solid lines. ***D***, Image of the sinusoidal fit amplitude $V_m$ at all points in space. ***E***, Image of the angular phase shift of the fit $\theta_B^{max}$ relative to $\theta_B = 0^o$ at all points in space. Marked points are the same as those in ***B***, corresponding to the $V_{ph}$ vs. $\theta_B$ data in ***C***.

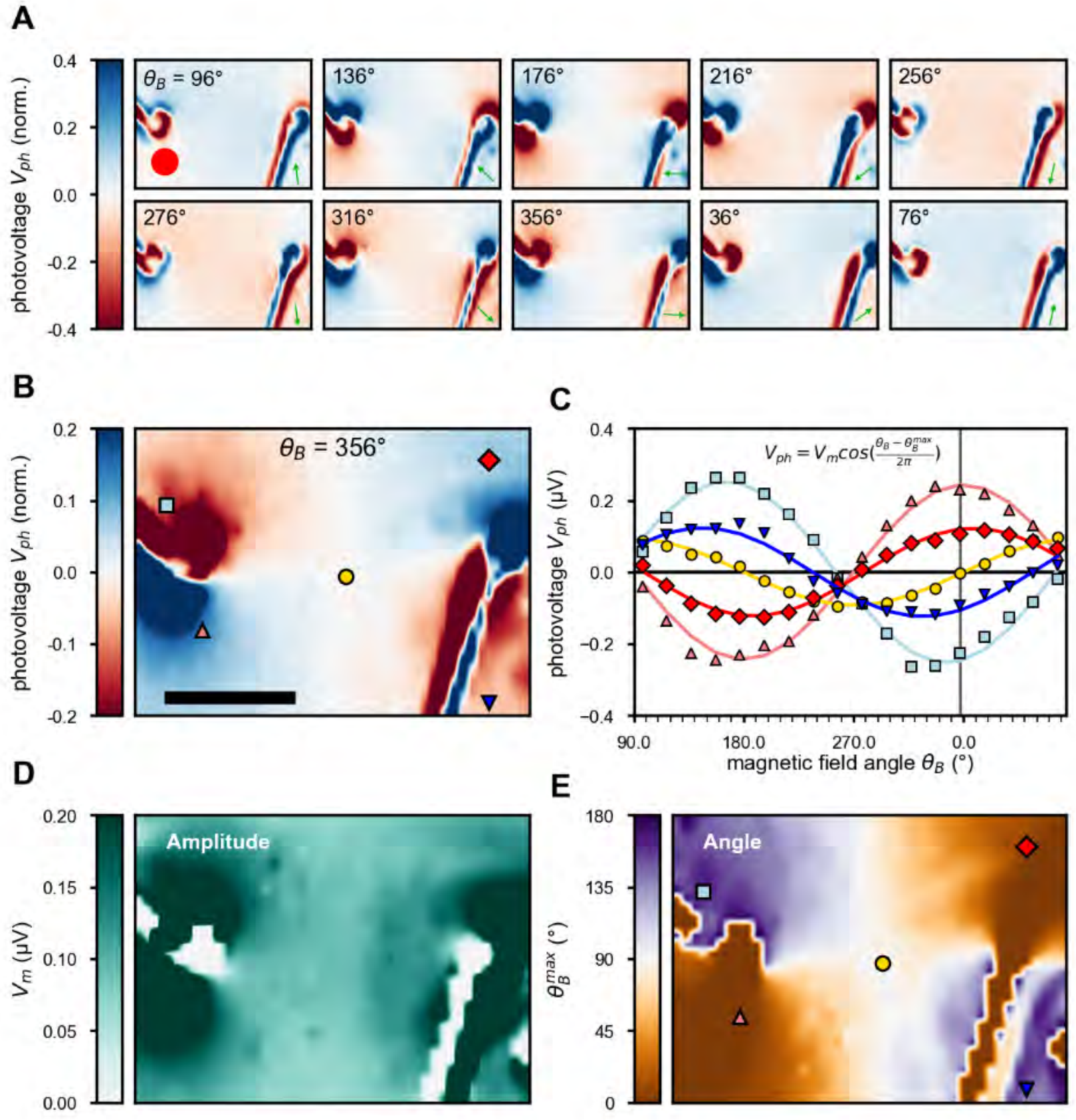

**Fig. S10.**

*Analyzing Field Dependent Effects in an un-patterned Pt/YIG Device.* Replica of main text Figure 2 using an un-patterned Pt/YIG device. ***A***, A subset of magnetic field-dependent photovoltage maps at 10 different magnetic field angles. $\theta_B$ is labelled top left corner and is indicated by green arrows. Red circle indicates the FWHM of the beam spot (FWHM = 27 µm). ***B***, Detailed view of the anomalous photovoltage features at $\theta_B = 357^{\circ}$. Scale bar 50 microns for all images. ***C***, $V_{ph}$ vs. $\theta_B$ for 5 points marked in ***B***. Data points share the same colors and shapes as in ***B***. At each point in space, $V_{ph}$ vs. $\theta_B$ is fit to the function $V_{ph}(\theta_B) = V_m \cos(\theta_B - \theta_B^{max})$, shown as corresponding solid lines. ***D***, Image of the sinusoidal fit amplitude $V_m$ at all points in space. ***E***, Image of the angular phase shift of the fit $\theta_B^{max}$ relative to $\theta_B = 0^{\circ}$ at all points in space. Marked points are the same as those in ***B***, corresponding to the $V_{ph}$ vs. $\theta_B$ data in ***C***.

are shown here to complement figure 4 of the main text and highlight the consistency of the analysis across various geometries. The top panel in figures S8-S9 correspond to a selection of 10 out of 18 available magnetic field orientations that can be seen for most devices in figure S7.

Having reduced the 3-dimensional data-cube for $V_{ph}$ to two different 2-dimensional maps for $V_m$ and $\theta_B^{max}$, we begin to study the anomalous spatial dependence in more detail. As the underlying response is field-dependent, it is reasonable to assume the $\theta_B^{max}(x,y)$ and $V_m(x,y)$ can be treated as components of a vector field. As a first step in analyzing the discrete data this way, we plot $V_m(x,y)$ and overlay it with a field of arrows pointing in the direction $\theta_B^{max}(x,y)$. The arrows are generated automatically from the phase data using the python graphical plotting function quiver from the matplotlib library. Figure 3A of the main text highlights the results of this method. We then use the python graphical interpolation function *streamplot* from the matplotlib library to interpolate between the arrows and produce a flow field through the device, shown in figure 3B and 4c of the main text. In a similar manner, we use the *streamplot* method to plot the interpolation of the field orthogonal to $\theta_B^{max}(x,y)$ at each point in space, *x* and *y*, a field explored by the main text and illustrated in figure 4D.

**S5. Discounting Competing Magnetic Effects**

As the spatial dependence of $V_{ph}$ was not as predicted from the ISHE alone, it is reasonable to ask if other magnetic effects could be present in the system. As discussed in section S4, the Seebeck effect data is calculated by averaging the photovoltage amplitudes for a single spatial point with opposite magnetic field sign – so the Seebeck effect data is still a data-cube with a 3rd dimension that has been cut in half. As can be seen in Figure S11, analyzing the standard deviation of amplitude for the Seebeck effect data across the different magnetic field orientations indicates that any additional effects (other than Seebeck effect or LSSE/ISHE) are negligible. Indeed, in this work, any additional field-dependent effects must lie well below the highly sensitive voltage noise floor of our measurement. As a control measurement, another Hall bar device was fabricated with only the platinum and GGG layers. As can be seen in Figure S12, only the conventional Seebeck Effect is observed.

The Hall effect, discovered by Edwin Hall in 1879[46], is a result of the Lorentz force on moving charges in a magnetic field and is characterized by a current deflection in the direction orthogonal to both the current and the external magnetic field. As this experiment relies heavily on the use of a Halbach array which produces a large magnetic field to saturate the magnetization of the YIG, the influence of this effect on the experimental results must be accounted for. The Lorentz force, $\vec{F} = q\vec{v} \times \vec{B}$, exerts a force on the charge carriers causing them to drift. For the duration of this experiment, $\vec{B}$ is purely in-plane and voltage probes are located on opposite edges of the devices, so measurable current also occurs in-plane. Thus, the Hall effect should generate some unmeasurable out-of-plane current. However, the Hall effect should also be present in the same devices missing the YIG layer. As the Pt/GGG signal, shown in figure S12 contains only the Seebeck effect and all other effects can account for less than 0.1% of the total signal, the Hall effect should also be negligible in our system. This same measurement also discounts the Nernst effect, in which $V = B \times T$. The anomalous versions of these effects, which rely on magnetization (*M*) rather than B, must also be accounted for. In a similar system with 5nm-thick Pt layers, it is known that the anomalous and ordinary Hall effects should be of comparable magnitude[38]. As we have shown the ordinary Hall effect to be negligible, it is assumed that the anomalous Hall effect is negligible as well. In the literature, several experiments have failed to detect the anomalous or proximity Nernst effects in similar systems[37]. While this lack of findings does not definitively rule out the possibility of interference with these effects, it indicates they are likely very small if present at all.

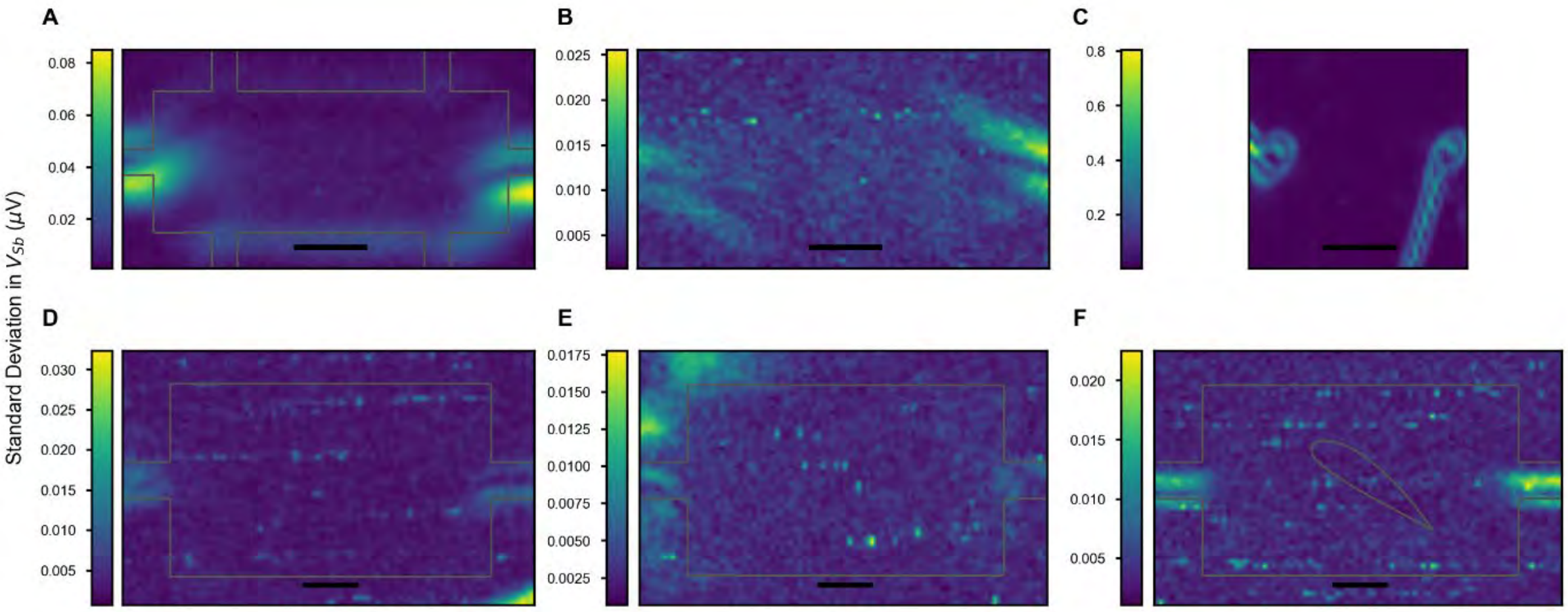

**Fig. S11.**

*Standard deviation in Seebeck extraction across scans taken at various magnetic field orientations.* Plots are separated by device geometry, ***A***, is the standard Hall Bar device used throughout the text, ***B***, is the terraced Hall cross, ***C***, is the un-patterned Pt/YIG device, ***D***, is the large, 2-contact device, ***E***, is the 0º electrofoil and ***F***, is the 35º electrofoil. Scale bars are each 50 microns long. Each spatial position on each device is measured 18 times, once every 20º. The Seebeck effect is extracted by averaging two measurements with opposite field angles, resulting in 9 individual results for the averaging process; panels ***A***-***F*** represent the standard deviation of these averages.

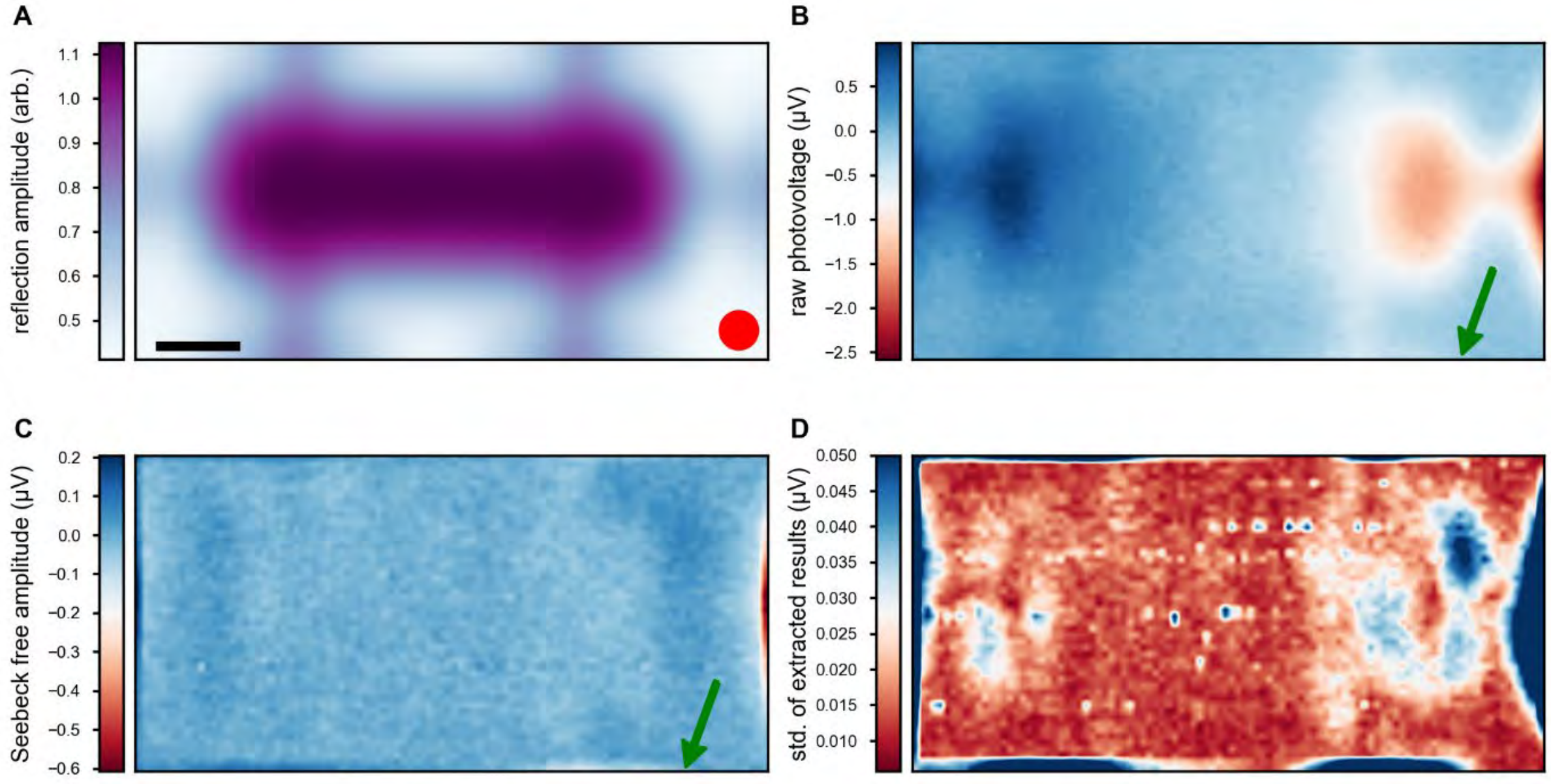

**Fig. S12.**

*Null Results of Pt-GGG measurement.* ***A***, A reflection image of the Pt-GGG Hall bar device. The device geometry is the same as the Hall bar used throughout the text. The beam spot is 27 microns FWHM and the scale bar is 50 microns long. ***B***, Raw photovoltage response for the Pt-GGG device. Black arrow represents the magnetic field angle when the photovoltage was recorded. ***C***, Pt-GGG measurement after Seebeck results have been subtracted, measured at the same angle as ***B***. ***D***, Standard deviation of the Seebeck-subtracted signal across all measured angles, indicating the results in ***C*** are representative of all field orientations.